\documentclass[conference]{IEEEtran}
\usepackage{cite}

\ifCLASSINFOpdf
\usepackage[pdftex]{graphicx}
\else
\fi
\usepackage{amsmath}
\usepackage{amsfonts} 
\usepackage{amsthm}
\usepackage{amssymb}
\usepackage[boxruled]{algorithm2e}
\usepackage{url}

\usepackage{bbm}

\usepackage{kbordermatrix}

\newcommand{\pfun}{\mathop{\hbox{$\to$\kern-7pt\raise.9pt\hbox{\scalebox{1}[.55]{$|$}}\kern4pt} }}

\usepackage{array}

\begin{document}

\title{Supercomputing Enabled Deployable Analytics \\ for Disaster Response}

\author{\IEEEauthorblockN{Kaira Samuel, Jeremy Kepner, Michael Jones, Lauren Milechin, Vijay Gadepally,  \\ William Arcand, David Bestor, William Bergeron, Chansup Byun,  Matthew Hubbell,\\  Michael Houle, Anna Klein, Victor Lopez, Julie Mullen, Andrew Prout, \\ Albert Reuther, Antonio Rosa, Sid Samsi, Charles Yee, Peter Michaleas
\\
\IEEEauthorblockA{MIT
}}}
\maketitle

\begin{abstract}
First responders and other forward deployed essential workers can benefit from advanced analytics.  Limited network access and software security requirements prevent the usage of standard cloud based microservice analytic platforms that are typically used in industry.  One solution is to precompute a wide range of analytics as files that can be used with standard preinstalled software that does not require network access or additional software and can run on a wide range of legacy hardware. In response to the COVID-19 pandemic, this approach was tested for providing geo-spatial census data to allow quick analysis of demographic data for better responding to emergencies.  These data were processed using the MIT SuperCloud to create several thousand Google Earth and Microsoft Excel files representative of many advanced analytics.  The fast mapping of census data using Google Earth and Microsoft Excel has the potential to give emergency responders a powerful tool to improve emergency preparedness.  Our approach displays relevant census data (total population, population under 15, population over 65, median age) per census block, sorted by county, through a Microsoft Excel spreadsheet (xlsx file) and Google Earth map (kml file). The spreadsheet interface includes features that allow users to convert between different longitude and latitude coordinate units. For the Google Earth files, a variety of absolute and relative colors maps of population density have been explored to provide an intuitive and meaningful interface.  Using several hundred cores on the MIT SuperCloud, new analytics can be generated in a few minutes.
\end{abstract}

%
\IEEEpeerreviewmaketitle

\section{Introduction}
\let\thefootnote\relax\footnotetext{
This material is based upon work supported by the Assistant Secretary of Defense for Research and Engineering under Air Force Contract No. FA8702-15-D-0001, National Science Foundation CCF-1533644, and United States Air Force Research Laboratory Cooperative Agreement Number FA8750-19-2-1000. Any opinions, findings, conclusions or recommendations expressed in this material are those of the author(s) and do not necessarily reflect the views of the Assistant Secretary of Defense for Research and Engineering, the National Science Foundation, or the United States Air Force. The U.S. Government is authorized to reproduce and distribute reprints for Government purposes notwithstanding any copyright notation herein.
}

\begin{figure}[htb]
  	\centering
    	\includegraphics[width=\columnwidth]{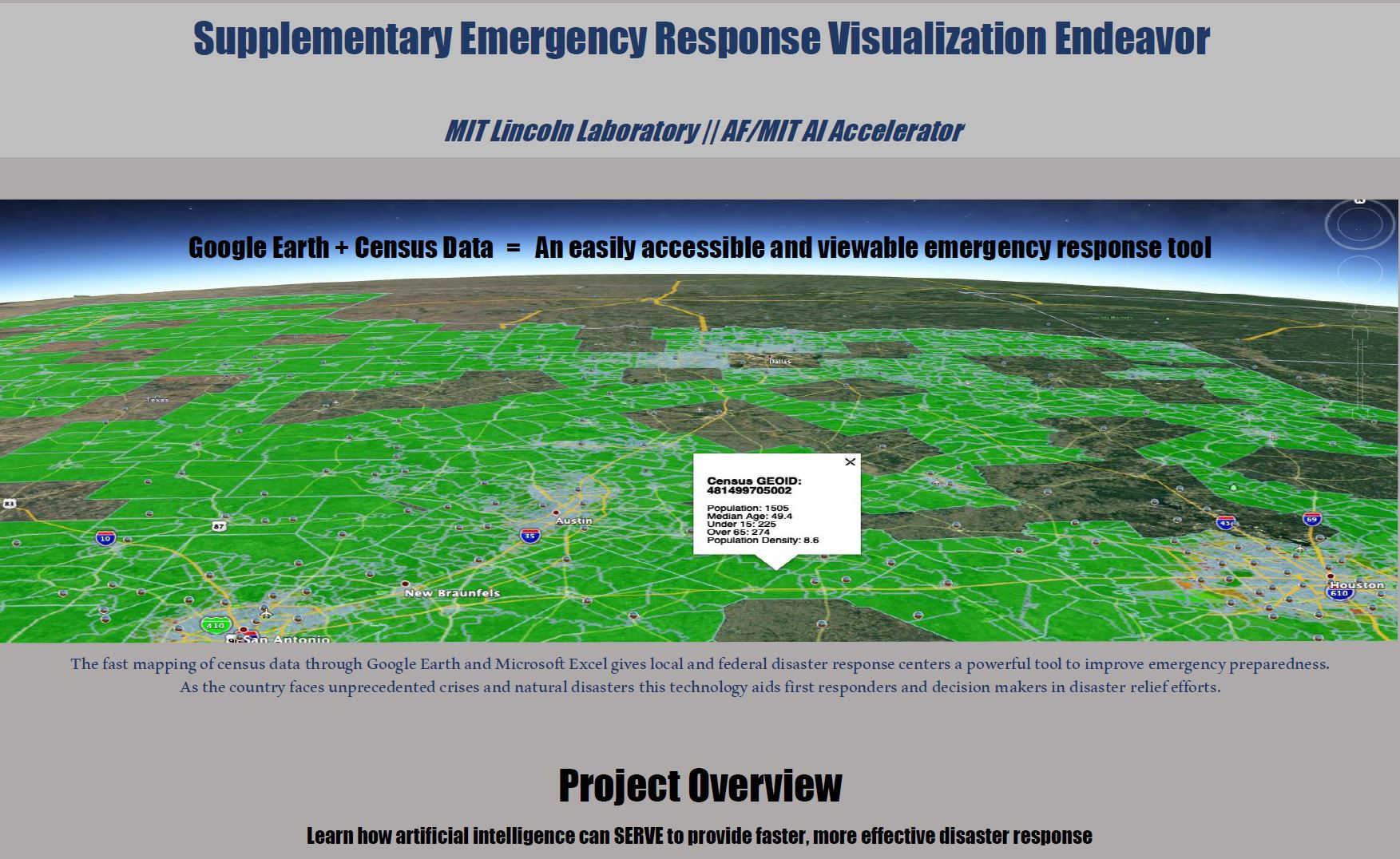}
	\caption{Above is the header of the one page description that has been used to publicize SERVE to potential users. Below the portion that is shown, there is a description of the tool and possible applications detailed.}
      	\label{fig:SERVEpage}
\end{figure}

Microservice-based cloud analytics have become common in many industries and large ecosystems exist to support these services \cite{villari2016osmotic, zschornig2017personal, khaleq2018cloud, ali2018design, li2019microservice, lu2020microservice}.  First responders and other forward deployed essential workers can benefit from advanced analytics, but limited network access and software security requirements often prevent the usage of these services \cite{matsuoka2019data}.  In contrast, prestaging analytics as files on emergency response systems is relatively simple as long as they can be read by standard pre-installed applications, such as Microsoft Excel and Google Earth.  Trading off computational power, network, and data is a standard practice in distributed systems design \cite{li2017fundamental, yan2018storage, nan2018dynamic, sun2019communications}.  The extreme case of reducing deployed computing and networking to near zero is achievable at the cost of pre-computing a large permutation space of possible analytics for a user to choose from, but may require significant computational resources.

In response to the COVID-19 pandemic, this pre-computed analytic approach was tested for providing geo-spatial census data to allow quick analysis of demographic data for better responding to emergencies.   Data mapping has been extrapolated to pandemic and other disaster relief to vastly improve the efficiency of emergency responses.  For many fast mapping processes, complex programs are often necessary for processing the data. While this can be beneficial for some computer systems, emergency response computers generally are not able to run these large programs.  Mapping census data provides abundant information to emergency respondents, including how dense the population is and how vulnerable the population is to a specific event. An MIT effort has been created to provide information on the technology, named SERVE (Supplementary Emergency Response Visualization Effort), as seen in Figure~\ref{fig:SERVEpage}, and has been provided to State Emergency Operation Centers (SEOCs).  The rest of this paper describes how SERVE was implemented as an example of the prestaged file-based approach for providing advanced analytics to emergency responders.

\section{Data}

Demographic boundary data can be any hierarchical representation of data from anywhere in the world.   The United States census boundary data consists of 56 states and territories, 3233 counties, and 219831 census block groups \cite{StateCounties,CensusBlocks} (which we refer to as ``census blocks'' for brevity).  The census data has been organized into a data structure to accelerate analytic construction (see \cite{FastMap} for additional details).  These census data in the data structure consist of names, polygon boundaries ($\bf{x}_{\rm poly}$, $\bf{y}_{\rm poly}$), bounding boxes ($\bf{x}_{\rm min}$, $\bf{x}_{\rm max}$, $\bf{y}_{\rm min}$, $\bf{y}_{\rm max}$), and Federal Information Processing Standards (FIPS) codes. At the top country level, this data structure is called {\tt us} and has the following fields containing the state FIPS, bounding box, polygon points, and a state structure for each of 56 state entities

{\footnotesize \begin{verbatim}
us.stateFP: [56x1 double]      
us.stateBB: [56x4 double]
us.stateXY: [56x1 struct]
us.state:   [1x56 struct]
\end{verbatim}}

\noindent For the Commonwealth of Massachusetts these values are

{\footnotesize \begin{verbatim}
us.stateFP(8)   = 25  
us.stateBB(8,:) =
   [-73.5081 -69.9284 41.2380 42.8866]
us.stateXY(8).X = [1x2612 double]
us.stateXY(8).Y = [1x2612 double]
\end{verbatim}}

The {\tt us.state} structure contains the corresponding county FIPS, bounding box, polygon points, and a county structure for each county in each state entity.  For the Commonwealth of Massachusetts there are 14 counties

{\footnotesize \begin{verbatim}
us.state(8).countyFP: [14x1 double]
us.state(8).countyBB: [14x4 double]
us.state(8).countyXY: [14x1 struct]
us.state(8).county:   [1x14 struct]
\end{verbatim}}

\noindent For Middlesex County in the Commonwealth of Massachusetts the values are

{\footnotesize \begin{verbatim}
us.state(8).countyFP(5)   = 17     
us.state(8).countyBB(5)   =
   [-71.8988 -71.0204 42.1568 42.7366]
us.state(8).countyXY(5).X = [1x508 double]
us.state(8).countyXY(5).Y = [1x508 double]
\end{verbatim}}

The {\tt us.state.county} structure contains the corresponding block FIPS, full FIPS, bounding box, and polygon points for each census block group in each county.  The county of Middlesex contains 1133 census block groups

{\footnotesize \begin{verbatim}
us.state(8).county(5).blockFP:   [1133x1 double]
us.state(8).county(5).blockFIPS: [1133x12 char]
us.state(8).county(5).blockBB:   [1133x4 double]
us.state(8).county(5).blockXY:   [1133x1 struct]
\end{verbatim}}

\noindent Finally, at the census block level, the additional population demographic information is included.  For one census block group at MIT in Middlesex County these values are

{\footnotesize \begin{verbatim}
us.state(8).county(5).blockFP(488)     = 3531012
us.state(8).county(5).blockFIPS(488,:) = 250173531012
us.state(8).county(5).blockBB(488,:)   =
   [-71.1021 -71.0908 42.3604 42.3660]
us.state(8).county(5).blockXY(488).X   = [1x74 double]
us.state(8).county(5).blockXY(488).Y   = [1x74 double]
us.state(8).county(5).blockPop(488)    = 1116
us.state(8).county(5).blockAge(488)    = 27.1
us.state(8).county(5).blockUnder15(488)= 120
us.state(8).county(5).blockOver65(488) = 45
us.state(8).county(5).blockPopSqMi(488)= 13950
\end{verbatim}}
The above structure can be readily computed from standard census data, does not increase the memory requirements of the data, and enables rapid construction of a wide range of analytics.

\section{Prestaged Analytics}

\begin{figure}[htb]
  	\centering
    	\includegraphics[width=0.415\columnwidth]{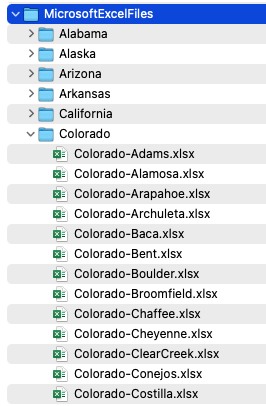}
    	\includegraphics[width=0.4\columnwidth]{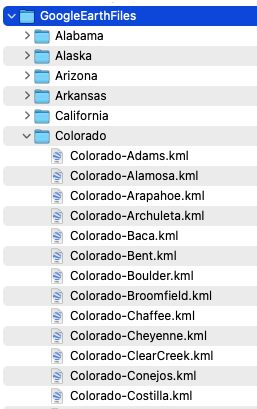}
	\caption{Directory structure of Microsoft Excel files and Google Earth files.  These allow the user to quickly locate census data and maps by state and county.}
      	\label{fig:AnalyticFiles}
\end{figure}

A primary aspect of prestaged analytics is to create files that precompute a wide range of permutations that can be readily navigated by users.  SERVE  consists of Microsoft Excel (xlsx) files and Google Earth Keyhole Markup Language (kml) files containing relevant geographic and demographic data. These files can be used simultaneously or separately depending on whether the user wants a visual component or an exclusively numerical representation of the census data.  The folder organization of these files reflect the natural geographic structure of the data (Figure~\ref{fig:AnalyticFiles}).

The code used to generate the xls and kml files is written in MATLAB. For the kml files, variables were created to represent the census data points. The major functions used to write the Google Earth files are fprintf to display the map and addPlacemark to add both the pop-up for each census block that displays the relevant census data and the population density colormap. For the xls files, variables were created to iterate through the census blocks of different counties within the census data structures. The major functions used to write the Excel files include writecell and writematrix, both of which provide the ability to write code using Excel functions while working with the MATLAB interface.

These several thousand Excel and Google Earth files are generated from the aforementioned data structure on the MIT SuperCloud. Using 10 40-core compute nodes (400 cores total) these analytic files can be generated in a few minutes, making it efficient to test different types of analytic files to determine those that are the most useful to the user.  The code was implemented using Matlab/Octave with the pMatlab parallel library \cite{Kepner2009}.  A typical run could be launched in a few seconds using the MIT SuperCloud triples-mode hierarchical launching system \cite{reuther2018interactive}.  Typical launch parameters were [10 5 8], corresponding to 10 nodes, 5 Matlab/Octave processes per node, and 8 OpenMP threads per process.  On each node, the 5 processes were pinned to 8 adjacent cores to minimize interprocess contention and maximize cache locality for the OpenMP threads \cite{byun2019optimizing}.  Each Matlab/Octave process was assigned a subset of the files to generate. Within each Matlab/Octave process, the underlying OpenMP parallelism was used on 8 cores to accelerate the processing.

\begin{figure}[htb]
  	\centering
    	\includegraphics[width=\columnwidth]{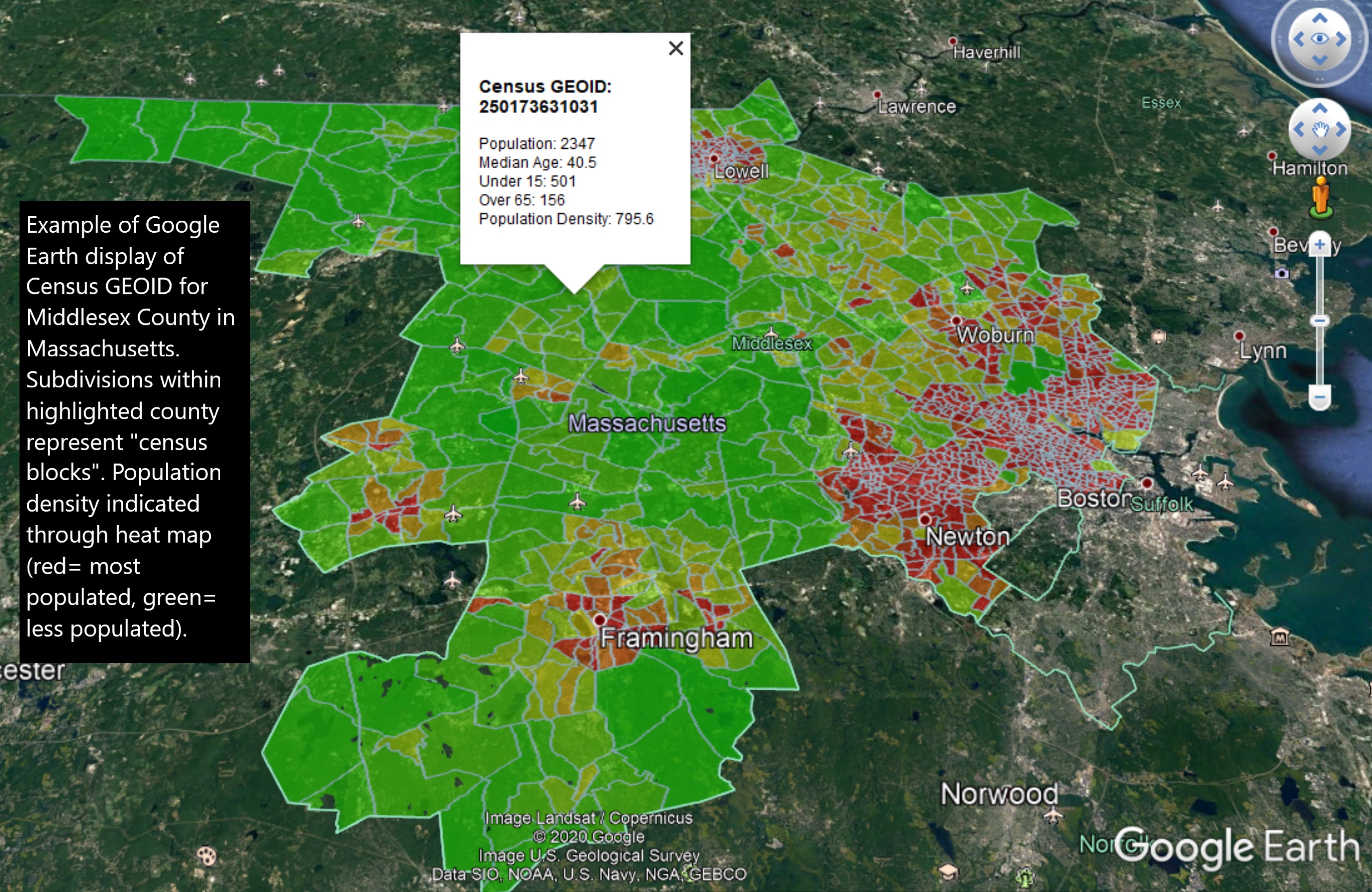}
	\caption{This figure shows a sample Google Earth file and a Census GEOID from one census block in Middlesec, Massachusetts.}
      	\label{fig:GoogleEarthEx}
\end{figure}

\subsection{Google Earth Files}
The Google Earth kml files, which can be imported into Google Earth, each correspond to specific United States counties. Each block has a Census GEOID that corresponds to the census block's relevant census data. Currently that includes the total population, population density, population over age 65, population under age 15, and median age. When the census block is clicked, the Census GEOID and corresponding data appears as a pop-up on the map, as seen in Figure~\ref{fig:GoogleEarthEx}.

\subsection{Microsoft Excel Files}
\begin{figure}[htb]
  	\centering
    	\includegraphics[width=\columnwidth]{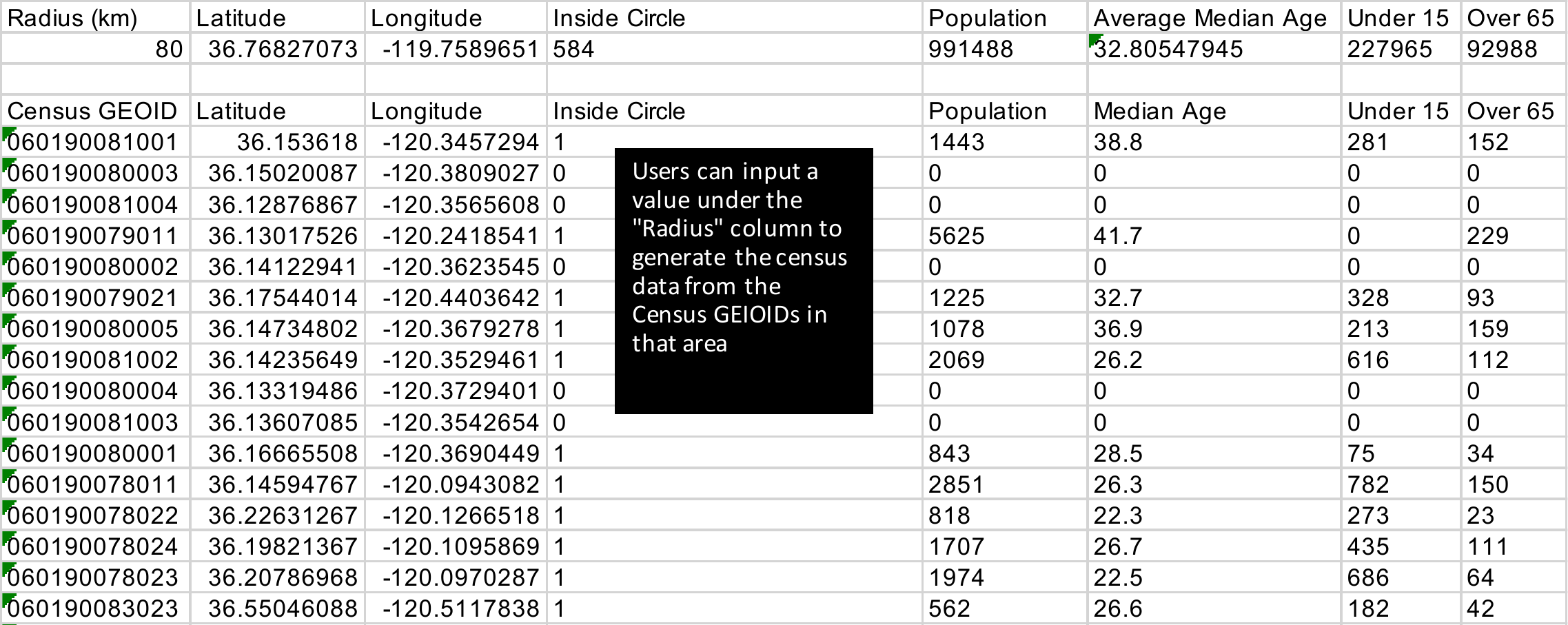}
	\caption{This figure shows a sample Microsoft Excel file from Fresno, California. The left most column contains Census GEOIDs for the county's census blocks and to their right lies all of the corresponding data.}
      	\label{fig:MicExcelEx}
\end{figure}

The Microsoft Excel files exhibit more detailed tallies of demographic data.  As in the kml files, each county is assigned a separate file containing several census blocks within it. The Excel files also contain the Census GEOIDs and relevant census data from the kml files, but can all be viewed at once as a list in the spreadsheet rather than as a pop-up. The Excel files also allows users to input an arbitrary radius around an affected county's central longitude and latitude coordinates. The tool then determines which census blocks are included in that radius and in turn provides the relevant census data within that desired area. An example of one of these Excel files can be seen in Figure~\ref{fig:MicExcelEx}.

\section{Dynamic Analytics}

To make the tools easier for people to use, some features were added to the original functionality of both the Microsoft Excel files and Google Earth files. These features utilize the ability of prestaged files to include more than just data.  As was shown in the Excel files, simple math formulas can be dynamically created and included in the Excel file that allow the user to dynamically work with the data.  These include formulas  providing users with more options for setting parameters in Excel and improving visualization in Google Earth.

\subsection{Excel Files}

\begin{figure}[htb]
  	\centering
    	\includegraphics[width=\columnwidth]{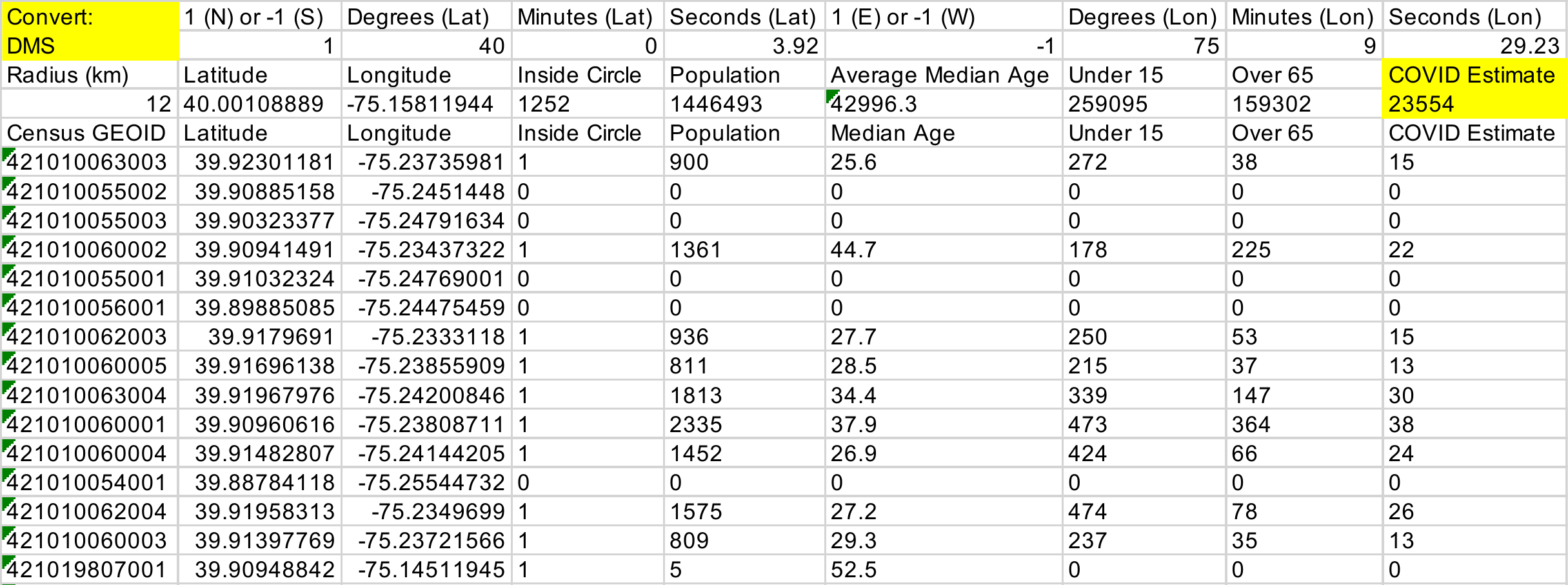}
	\caption{This figure of data from Philadelphia, Pennsylvania shows how a user can convert DMS coordinates into decimal degrees easily at the top of the spreadsheet. They can write DMS under the 'Convert' header and populate the rest of the cells following the headers in the first row. The longitude and latitude coordinates will then be converted into decimal degrees in the spreadsheet below.  The final column further shows how an estimate of the COVID cases could also be included.}
      	\label{fig:CoordConv}
\end{figure}

In first responder, government, and military operations, it is common to also use the degrees minutes seconds (DMS) and degrees minutes (DM) coordinate systems over the standard decimal degrees. The Excel files default units are decimal longitude and latitude, so a user would have to convert the coordinates externally if they had a different system. A header was therefore added to the top of the Excel files which would allow users to input DMS and DM coordinates and convert them immediately in Excel. The highlighted portions of Figure~\ref{fig:CoordConv} shows what this top section looks like when the file is opened and the user has DMS coordinates. In the future, more conversion capabilities can be added to the files depending on the users' needs. 

\subsection{Google Earth Files}

\begin{figure}[htb]
  	\centering
    	\includegraphics[width=\columnwidth]{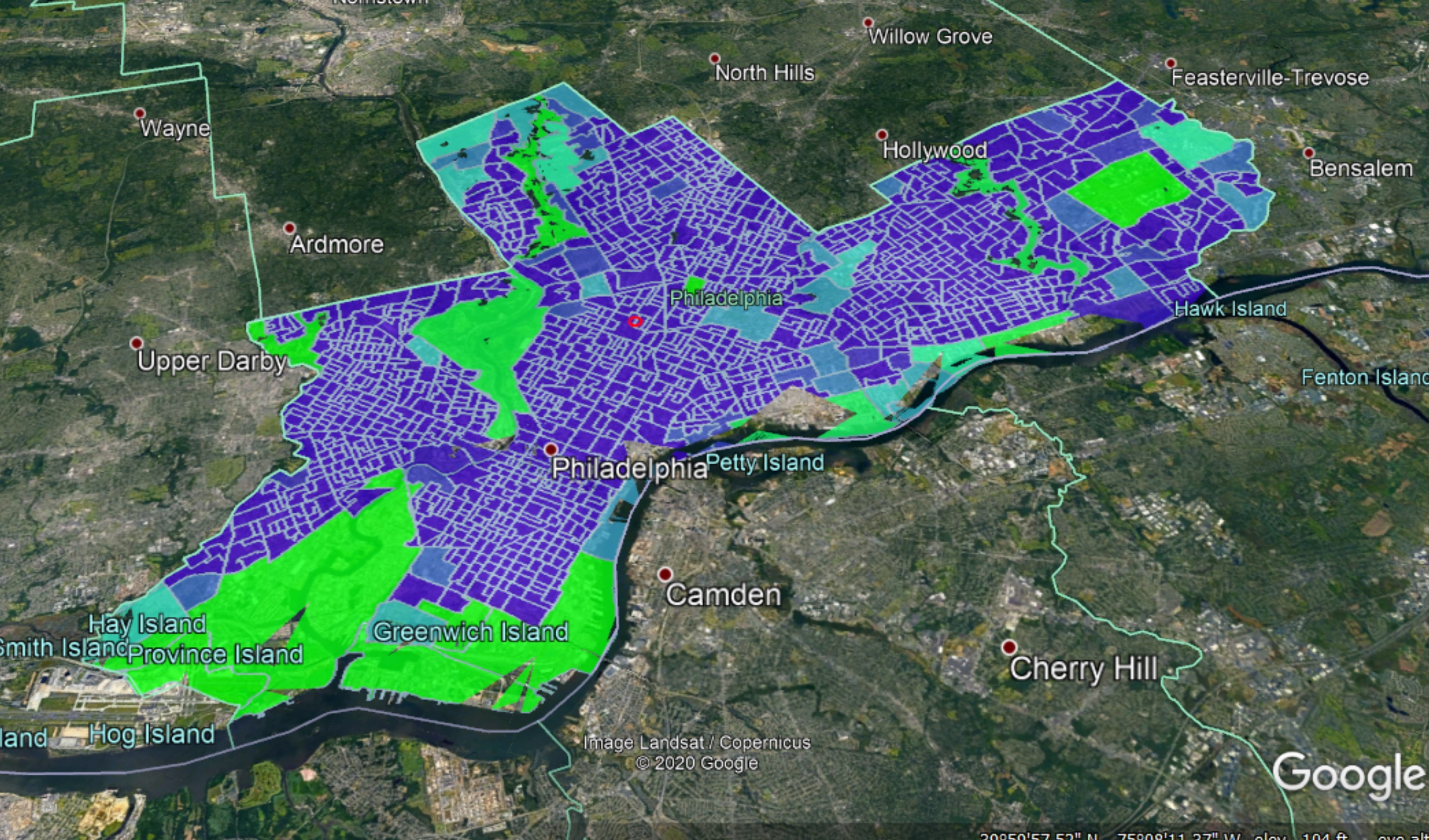}
	\caption{The above color map features red - blue shading for Philadelphia County.}
      	\label{fig:PDmap1}
\end{figure}
\begin{figure}[htb]
  	\centering
    	\includegraphics[width=\columnwidth]{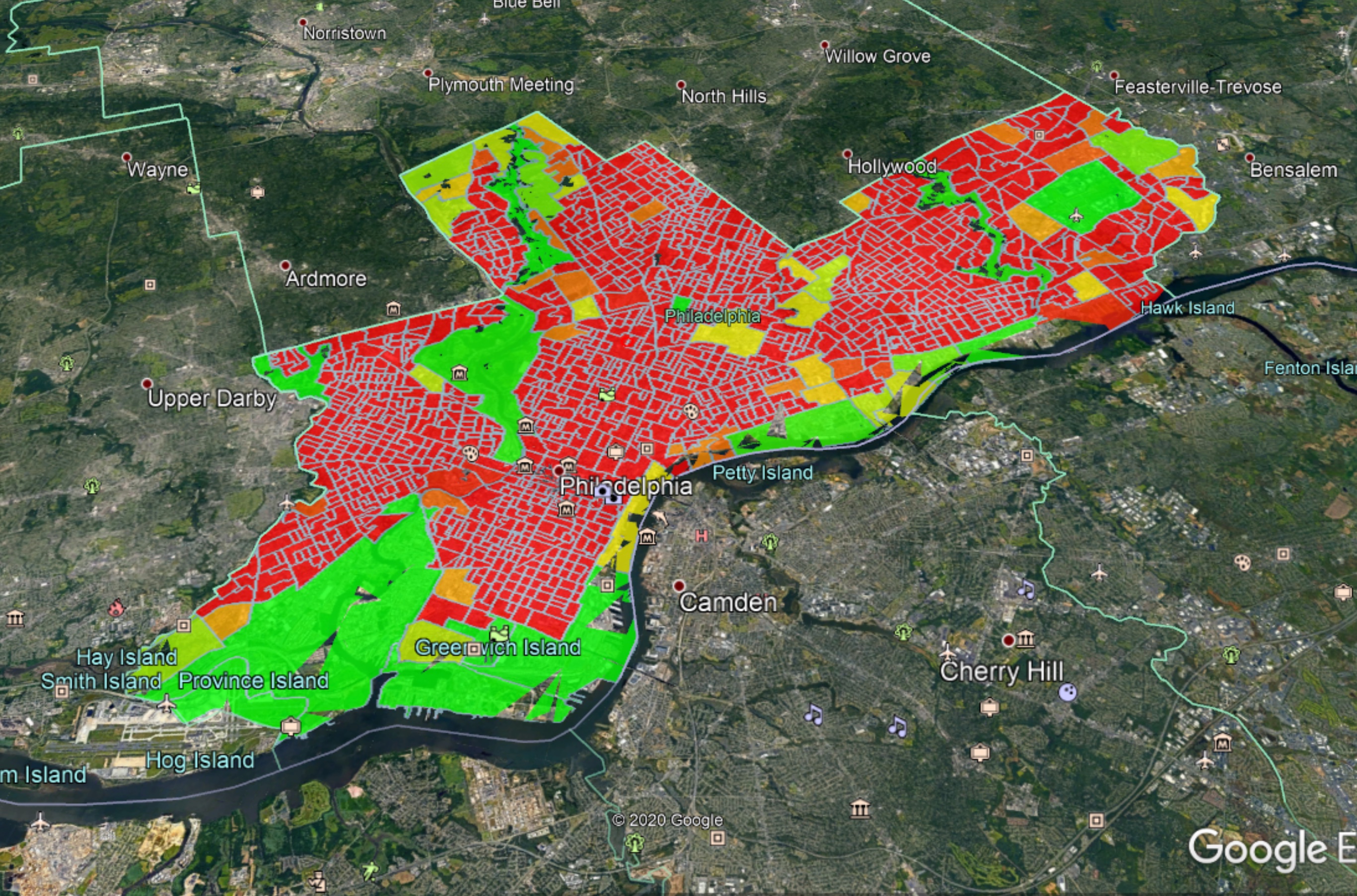}
	\caption{The above color map features green - red shading for Philadelphia County.}
      	\label{fig:PDmap2}
\end{figure}
\begin{figure}[htb]
  	\centering
    	\includegraphics[width=\columnwidth]{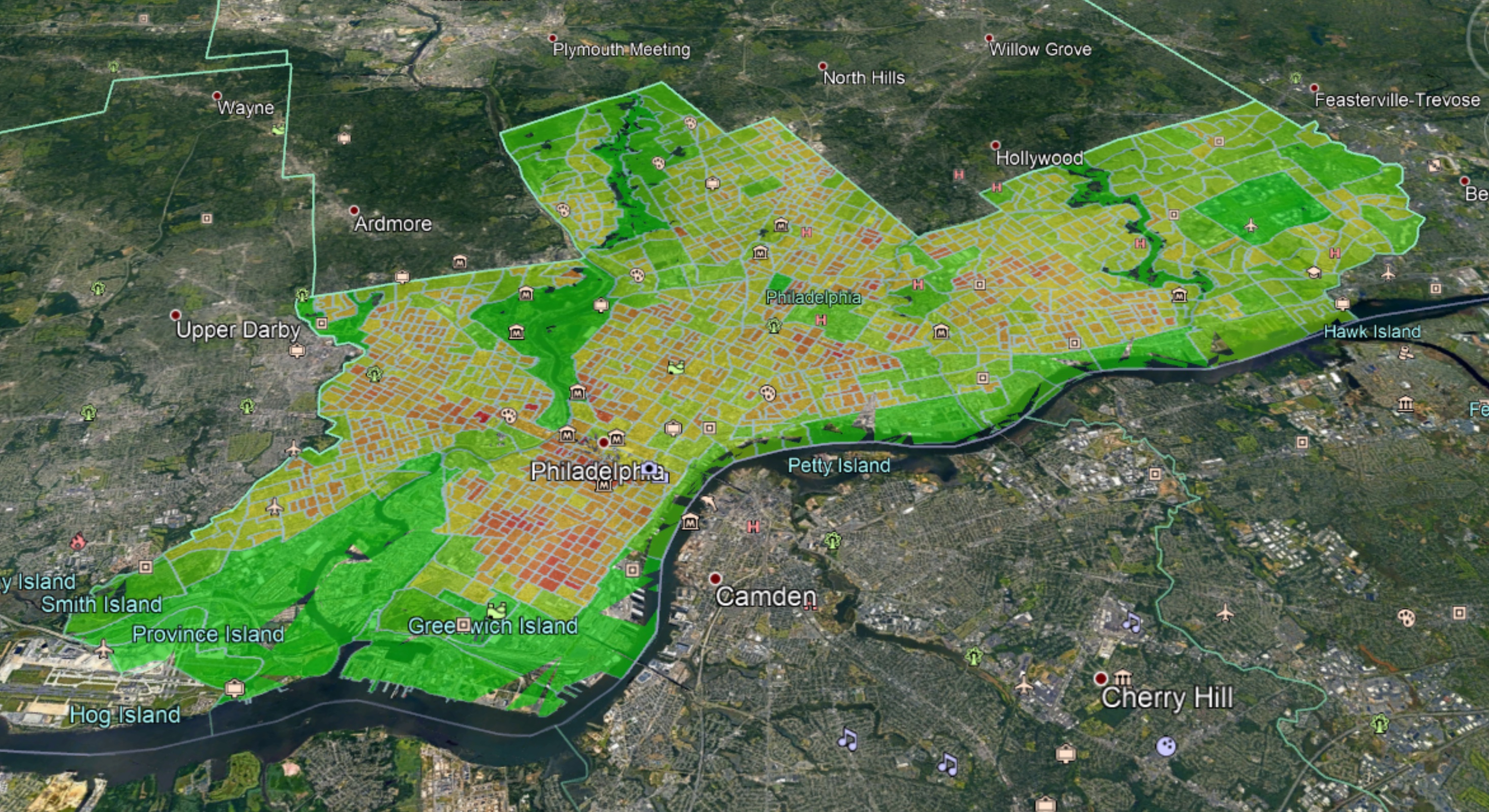}
	\caption{The above color map features a duller version of the green - red shading for Philadelphia County.}
      	\label{fig:PDmap3}
\end{figure}
\begin{figure}[htb]
  	\centering
    	\includegraphics[width=\columnwidth]{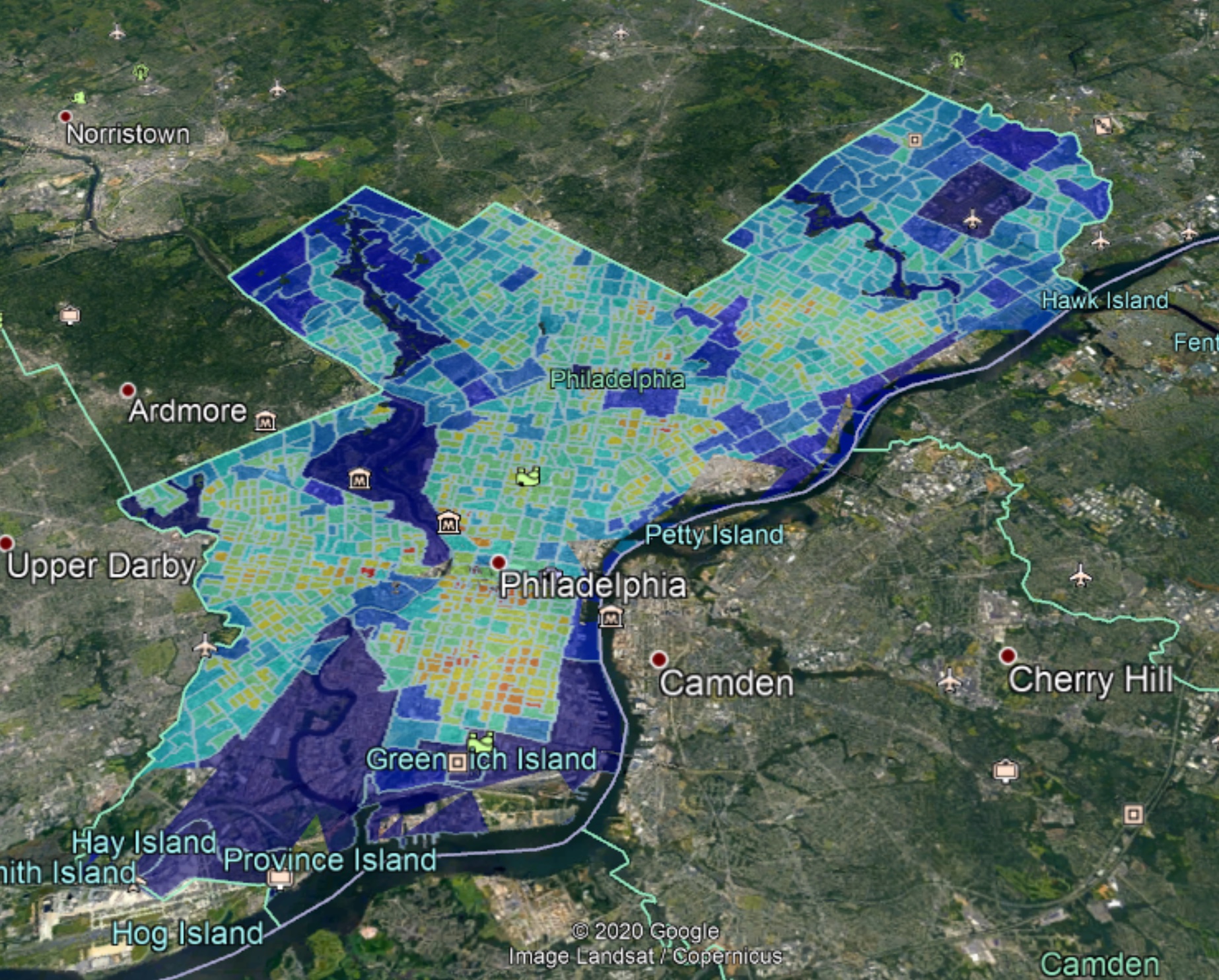}
	\caption{This figure features the Matlab built in 'jet' color map for Philadelphia County. The shading for this file has been done relative to just Philadelphia's census blocks.}
      	\label{fig:PDjetrel}
\end{figure}

The Google Earth files feature population density colored maps and it is important for the color scheme to properly convey the densities in a visually recognizable way. Selecting the color map involved iterating through a variety of color schemes. Figure~\ref{fig:PDmap1}, Figure~\ref{fig:PDmap2}, and Figure~\ref{fig:PDmap3} show some of the schemes that were tested.  In many cases the Matlab color scheme 'jet' was a popular choice for color maps. Figure~\ref{fig:PDjetrel} shows the jet color map from one of the kml files.

\begin{figure}[htb]
  	\centering
    	\includegraphics[width=\columnwidth]{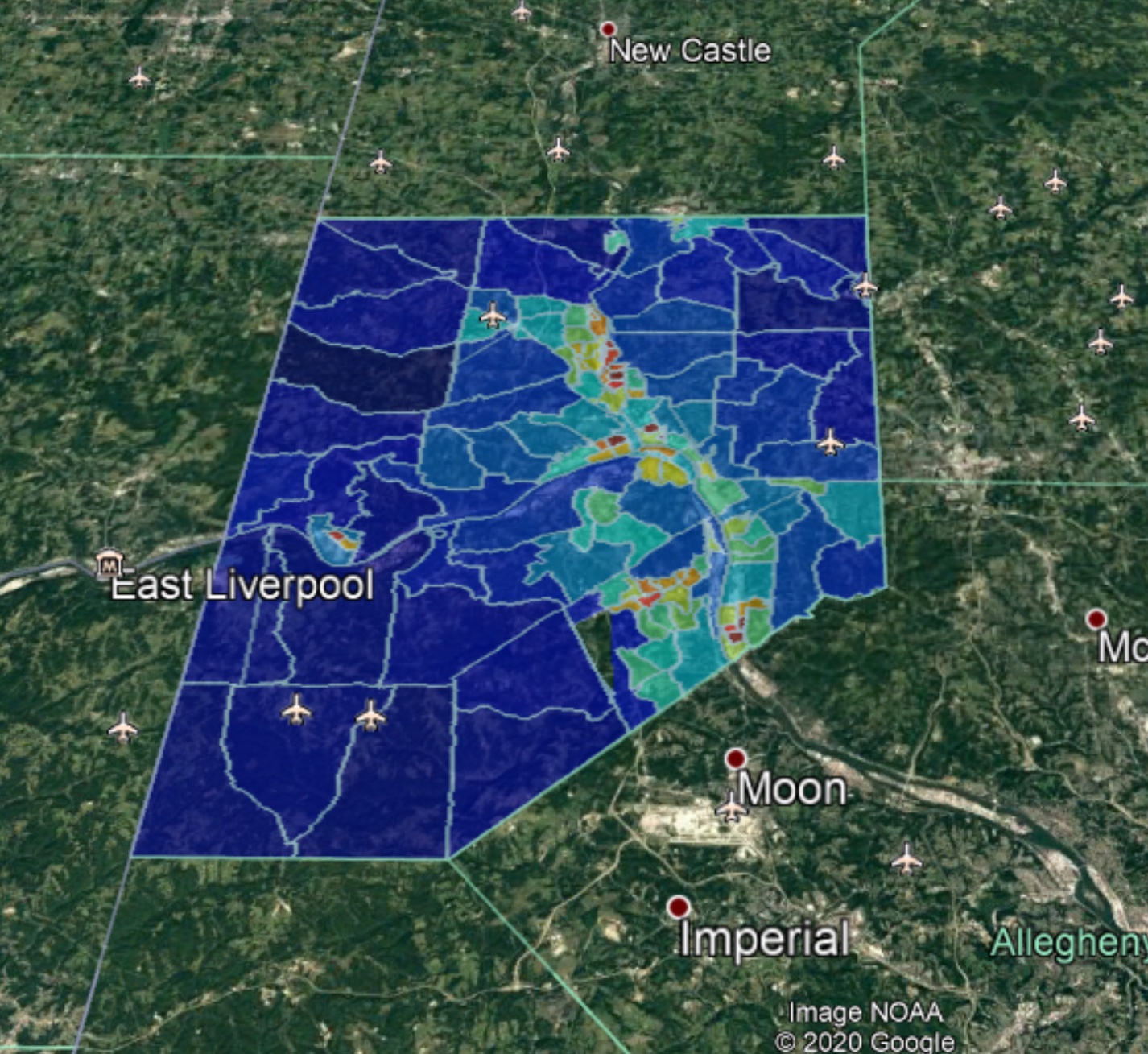}
	\caption{This figure features the 'jet' color map for Beaver County, Pennsylvania, where the shading has been done relative to only Beaver's census blocks. Although Beaver has a low population and low population density, there are red spots in certain census blocks that indicate greater density than other blocks in the county.}
      	\label{fig:Beaverjetrel}
\end{figure}

\begin{figure}[htb]
  	\centering
    	\includegraphics[width=\columnwidth]{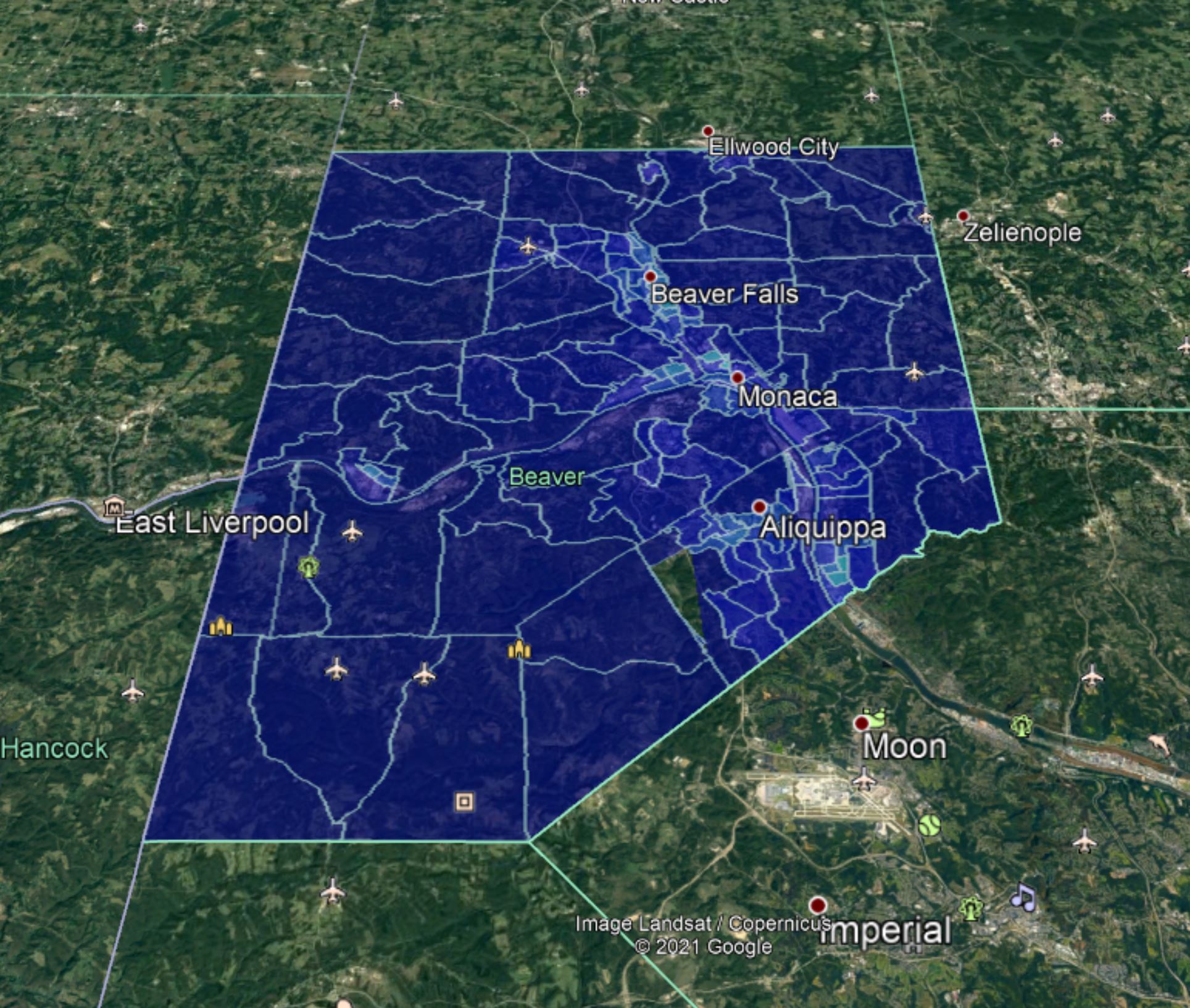}
	\caption{This figure features the 'jet' color map for Beaver County, Pennsylvania, where the shading has been done relative to the entire country. This configuration no virtually no warm colors, demonstrating the overall low population.}
      	\label{fig:Beaverjetabs}
\end{figure}

Population density normalized to the entire country as well as normalized within a county are both important views that can be used by emergency responders in different contexts. With COVID-19 it is important to look at the country level density comparisons because the pandemic is affecting everyone. However, with more localized disasters, such as fire or hurricane relief, it could be important to look at the population density distribution in singular affected counties. Two data sets were therefore compiled- one labeled 'relative' where the map is colored relative to the independent county densities, and one labeled 'absolute' where the map is colored relative to the entire country. A comparison can be seen in Figure~\ref{fig:Beaverjetrel} and Figure~\ref{fig:Beaverjetabs}.  The rapid turnaround afforded by supercomputing allows these analytics to be quickly constructed and tested.


\section{Summary}

Pandemics and other disasters pose an important problem that can be improved through the use of efficient data mapping. There is a potential for great impact by  processing large amounts of data without requiring complicated programs that are difficult to maintain or that require network access in the field. Emergency response computers generally are not able to run complicated programs, but they do have the capacity to support data files. Microsoft Excel and Google Earth are both widely used and well supported on emergency response computing systems. By providing emergency responders with ample data on population density and vulnerable populations of affected areas, ready access to mapping census data can improve the efficiency and efficacy of their response. 


\section*{Acknowledgments}
%
%

The authors wish to acknowledge the following individuals for their contributions and support: Bob Bond, Ronisha Carter, David Clark, Alan Edelman, Nathan Frey, Jeff Gottschalk, Tucker Hamilton, Chris Hill, Hayden Jananthan, Mike Kanaan, Tim Kraska, Charles Leiserson, Dave Martinez, Mimi McClure, Joseph McDonald, Christian Prothmann, John Radovan, Steve Rejto, Daniela Rus, Matthew Weiss, Marc Zissman.



\bibliographystyle{ieeetr}
\bibliography{DeployableAnalytics}
%

\appendices
\end{document}